\documentstyle[11pt,newpasp,twoside,epsf]{article}
\def\edcomment#1{\iffalse\marginpar{\raggedright\sl#1\/}\else\relax\fi}
\reversemarginpar

\def\Ha{H$\alpha$\ }

\begin{document}
\title{Molecular gas and star formation in BIMA SONG bars}
\author{Kartik Sheth}
\affil{Department of Astronomy, U. Maryland, College Park, MD 20742-2421}
\author{S.N. Vogel, A.I. Harris (U. Maryland), M.W. Regan (STScI), M.D. Thornley (NRAO), T.T. Helfer, T. Wong, L. Blitz, D.C-J. Bock (U.C.-Berkeley)}

\begin{abstract}
Using a sample of 7 barred spirals from the BIMA Survey of Nearby
Galaxies (SONG), we compare the molecular gas distribution in the bar,
to recent massive star formation activity.  In all 7 galaxies, \Ha is
offset azimuthally from the CO on the downstream side.  The maximum
offset, at the bar ends, ranges from 170-570 pc, with an average of
320$\pm$120 pc.  We discuss whether the observed offsets are
consistent with the description of gas flows in bars provided by the
two main classes of models: n-body models and hydrodynamic models.
This work\footnote{The entire poster at http://bima.astro.umd.edu/projects/bimasong/pubs/romeposter.ps} is
supported by NSF grants AST 9981289, AST 9981308.
\end{abstract}

\noindent{\bf I. Star Formation in Bars:} How and why do stars form where
they do ? To answer this question one may begin by studying star formation
activity and its molecular gas environment in a variety of
environments, such as bars, spiral arms, and rings.  Then by drawing
together physical processes and triggers that are common in these
different regions, we may shed light on how star formation
might occur.  Here we concentrate on bars in 7 galaxies from
the BIMA Survey of Nearby Galaxies (SONG).  We compare the molecular
gas distribution, traced by the CO (J=1-0) emission line, to the star
formation activity traced by the \Ha emission line.  While overlays
of CO and \Ha maps show qualitative agreement, we use a
cross-correlation analysis to quantify the relationship between the
two. \\

\noindent {\bf II. The Cross-Correlation Process:}
After matching the images in resolution, we deproject them and convert
them to polar coordinates.  To cross-correlate the images, we overlay
and shift one image against the other, at each step multiplying the
images pixel by pixel and recording the sum.  The offset lag at which
the two images match best has the largest correlation value.  The 
results of our analysis are shown in Figure 1. \\

\noindent{\bf III. Results: Interpreting in Context of Gas Flow Models:}
\vspace{-0.05in}
\begin{itemize}
\item In ALL 7 galaxies, \Ha is predominantly offset azimuthally
from the CO.  Moreover, the offset is always on the downstream side.
\item Maximum offset (measured at bar ends): 170--570 pc, Mean: 320$\pm$120 pc.
\end{itemize}

\begin{figure}
\plotone{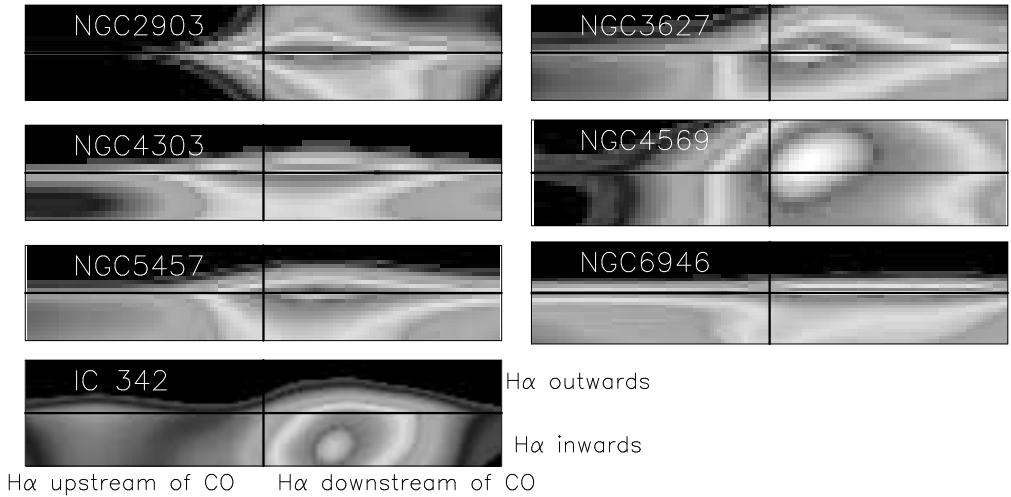}
\caption{\Ha-CO cross-correlation for all 7 bars. Above (Below) the
cross-hair indicates a radial offset of \Ha inwards (outwards) from
the CO.  Left (Right) of the cross-hair indicates an azimuthal offset
with \Ha upstream (or downstream) of CO.}
\end{figure}

\noindent {\bf IIIa. N-body simulations:} Gas clouds behave as in spiral
arms, crowding in the dust lane, but eventually crossing it on the
leading side (e.g., Combes \& Gerin 1985).  These models predict
the observed offset of CO and \Ha but fail to reproduce
the straight dust lanes and high shear across the dust lanes.
So these models may not be applicable in all cases. \\

\noindent {\bf IIIb. Hydrodynamic simulations:} Gas undergoes a
hydrodynamic shock at the dust lane, and flows down the dust lane,
never crossing the dust lane.  Though the dust lane is inhospitable
for star formation (e.g., Reynaud \& Downes 1998), stars may form in
dust spurs upstream of dust lane (Sheth et al. 2000).  Since none
of the 7 galaxies show offsets $>$ 500pc, star formation in these bars
may still occur in dust spurs because the distance between the spurs
and HII regions is consistent with typical gas speeds (50 km/s) and HII
region lifetimes.  But the offset also indicates that HII regions
spend most of their time on the leading side of the dust lane, thus
constraining star formation to occur very close (100-200 pc) to the
dust lane.  The dust lane may be involved, if not in triggering the
star formation, in formation of dust spurs. \\

\noindent {\bf IV. Conclusions:} \Ha emission is {\it always} offset
downstream from bar dust lanes; the offset is consistent with both
sets of gas flow models. \\
 
\noindent{\bf References} \\
\noindent Combes, F., \& Gerin, M. 1985, \aap, 150, 327 \\
\noindent Reynaud, D., \& Downes, D. 1998, \aap, 337, 671 \\
\noindent Sheth, K., Regan, M.W., Vogel, S.N. \& Teuben, P.J. 2000, \apj, 533, 221 \\
\end{document}